\documentclass{PoS}
\title{Intrinsic GeV-TeV gamma-ray emission from EHSP blazars}

\ShortTitle{$\gamma$-ray emission from EHSP blazars}

\author{{K. K. Singh}\thanks{Corresponding Author}\\
        Department of Physics, University of the Free State, Bloemfontein 9300, South Africa\\
        Astrophysical Sciences Division, Bhabha Atomic Research Centre, Mumbai 400085, India
        E-mail: \email{kksastro@barc.gov.in}}

\author{P. J. Meintjes\\
Department of Physics, University of the Free State, Bloemfontein 9300, South Africa\\
}

\author{N. Bhatt\\
Astrophysical Sciences Division, Bhabha Atomic Research Centre, Mumbai 400085, India\\
}

\author{ B. van Soelen\\
Department of Physics, University of the Free State, Bloemfontein 9300, South Africa\\
}

\abstract{Extremely High Synchrotron Peak (EHSP) blazars are observed to form a
small population of sources with high energy hump peaking at TeV
energies in their broad-band spectral energy distributions. The observed
gamma-ray emission from these sources at GeV-TeV energies is described
by unusual hard spectral indices. The observed spectral and temporal
characteristics of these sources challenge the standard leptonic models
for the broad-band emissions from blazars. Therefore, such sources
provide astrophysical sites to investigate directly the particle
acceleration, cooling of relativistic particles and indirectly probe the
cosmological quantities like extragalactic background light (EBL) and
intergalactic magnetic field (IMF) in the Universe. In this study, we
investigate the spectral properties of the gamma-ray emission from EHSP
blazars using observations from the Fermi-LAT catalogues (3FGL and
3FHL) along with the TeV observations using ground based telescopes.
The observed TeV gamma-ray spectra are corrected for the EBL
absorption using the most recent and updated EBL models to determine
the intrinsic spectrum at the source. The intrinsic TeV spectra are combined
with the MeV-GeV observations from the Fermi-LAT to study the gamma-
ray emission from EHSP blazars in the broad energy band. The intrinsic
gamma-ray spectra are then used to estimate the position of high energy
peak in the spectral energy distribution. We also present the qualitative
description for the observed spectral properties of EHSP blazars using
different physical scenarios and discuss their importance for the
upcoming CTA observatory.}
\FullConference{36th International Cosmic Ray Conference -ICRC2019-\\
		July 24th - August 1st, 2019\\
		Madison, WI, U.S.A.}

\begin{document}

\section{Introduction}
Blazars are a dominant class of radio-loud active galactic nuclei (AGN) which emit non-thermal radiation over 
the entire electromagnetic spectrum from radio to very high energy (VHE; E $>$ 50 GeV) gamma rays. The observed 
broadband radiation from blazars is interpreted as the multi-wavelength emission from the relativistic plsama jet 
which points in the direction of the observer at small viewing angle \cite{Padovani2017}. The spectral energy 
distribution (SED) of blazars is described by two characteristic humps peaking at low (IR/optical/X-ray) and 
high ($\gamma$-ray) energies respectively. The physics of the low energy hump in the blazar SED is well 
established and is attributed to the synchrotron emission from the relativistic leptons (mostly electrons and positrons) in 
the jet magnetic field. On the basis of the position of rest frame synchrotron peak frequency ($\nu_{peak}^{sync}$) in the SED, 
blazars are classified as low synchrotron peaked (LSP: $\nu_{peak}^{sync}~<~10^{14}$ Hz), intermediate synchrotron peaked 
(ISP: $10^{14}$ Hz$~<~\nu_{peak}^{sync}~<~10^{15}$ Hz), high synchrotron peaked (HSP: $10^{15}$ Hz$~<~\nu_{peak}^{sync}~<~10^{17}$ Hz) 
and extremely high synchrotron peaked (EHSP: $\nu_{peak}^{sync}~>~10^{17}$ Hz) blazars \cite{Costamante2001,Abdo2010}. 
The mechanism of $\gamma$-ray emission and origin of high energy (HE; E $>$ 0.1 GeV) hump in the SED are still under debate 
in blazar research. In general, $\gamma$-ray photons at GeV-TeV energies are supposed to be produced by the inverse Compton (IC) 
scattering of low energy photons by the relativsitic leptons in the blazar jet under the frame-work of leptonic models 
\cite{Bottcher2007,Agudo2011}. Alternatively, in a hadronic scenario, proton synchrotron and photo-hadronic interactions are also 
invoked to explain the observed $\gamma$-ray emission from blazars \cite{Aharonian2002,Bottcher2013}. Leptonic and Lepto-hadronic 
(Hybrid) models have been successfully used to describe the VHE $\gamma$-ray emission from most of the blazars observed so far. 
Variability in the spectrum and light curves in almost all energy bands at different timescales ranging from few minutes to years 
is observed to be a common feature of blazars \cite{Albert2007,Singh2018}. However, EHSP sources challenge the $\gamma$-ray emission 
models proposed for blazars and lack rapid flux variability in $\gamma$-ray energy band \cite{Singh2019}. In this study, we use a 
sample of four well known EHSP blazars to understand the $\gamma$-ray emission processes from these peculiar sources. 
\section{VHE $\gamma$-ray Emission from EHSP Blazars}
The intrinsic VHE $\gamma$-ray emission observed from EHSP blazars is described by a power law with hard photon spectral 
index $\Gamma_{int}~\le~$2. Due to their hard spectra, the peak of $\gamma$-ray component in the broadband SED is located 
at energies above 1 TeV. This challenges the standard leptonic emission models based on IC scattering of 
synchrotron photons or external photons for $\gamma$-ray emission from the blazars. The IC scattering in the Thomson 
regime decreases with increasing electron energy due to the lower energy density of target photons in 
the emission region of the jet. In the Klien-Nishina (KN) regime, the spectrum of VHE $\gamma$-rays is steeper or soft 
due to a decrease in the efficiency of the IC scattering \cite{Singh2019}. The hard $\gamma$-ray spectra also indicate 
injection of relativistic particles having very hard spectrum with index less than 1.5 in the emission region of the jet. 
This poses a strong challenge to the accelearion of particles through diffusive shock acceleration in the jet. Therefore,
following alternative scenarios including hadronic processes have been invoked to explain the gamma-ray emission from 
the EHSP blazars:
\begin{itemize}
	\item IC scattering of synchrotron photons by the relativsitic electrons in the emission region under the 
	      frame-work of a simple one zone leptonic model can produce VHE $\gamma$-rays with extremely hard spectra 
	      assuming large Doppler factor values as well as very high values of minimum Lorentz factor of the electrons 
	      with low radiative efficiency, and physical conditions far way from the equipartion \cite{Costamante2018}. 
	     Large Doppler factors indicate emission region moving relativistically at small viewing angle which is not 
		consistent with radio observations of the knots in the jet. 
         
	\item IC scattering of low energy seed photons by a narrow energetic particle distribution (relativistic Maxwellian 
		produced by a stochastic \emph{Fermi}-second order acceleration) in the Thomson regime can emit hard 
	      TeV spectra \cite{Lefa2011}. Also within the frame-work of single zone leptonic models, a power law 
	     distribution with a low energy cutoff of electrons in VHE band can in principle produce hard $\gamma$-ray spectra.

     \item  IC scattering of cosmic microwave background (CMB) photons by the ultra-relativistic electrons produced by the 
	     shock acceleration in an extended jet (kpc-scale) can result in the slowly variable VHE $\gamma$-ray emission 
	     with very hard spectrum in the Thomson limit \cite{Bottcher2008}. This scenario suggests separate components 
	     for X-ray and $\gamma$-ray emissions from the blazar.	
	 
      \item Internal $\gamma-\gamma$ absorption of VHE $\gamma$-rays due to their interaction with dense narrow-band 
	    low energy radiation fields close to the emission region in the jet can lead to the formation of TeV spectra 
	    with arbitrary hardness in a natural way \cite{Aharonian2008}. This does not require any modification in the 
	    particle spectra or particle acceleration models.	

      \item Lepto-hadronic processes including IC scattering of synchrotron photons produced by leptons, proton synchrotron and 
	     synchrotron emission from the cascade initiated by the proton-$\gamma$ interaction can collectively produce 
	     hard $\gamma$-ray spectra from blazars \cite{Cerruti2015}.	

      \item In a completely different scenario for EHSP blazars, $\gamma$-rays are not produced in the jet, instead in the 
	      intergalactic space. The ultra-high energy cosmic rays (UHECRs) escaping from the jet (mostly protons with 
	      energy above 10$^{18}$ eV) and beamed towards the observer due to weak cosmic magnetic field, interact with 
	      the CMB or UV/IR/optical photons through photo-meson and pair production in the intergalactic medium and 
		initiate development of electromagnetic cascade. The $\gamma$-rays with hard spectra are produced by the 
		secondary particles in the cascade \cite{Essey2011}. Recent detection of high energy astrophysical 
        	neutrinos from a few blazars support this hypothesis for VHE $\gamma$-ray production.    	
\end{itemize}
Therefore, the EHSP blazars with hard VHE $\gamma$-ray spectra provide a unique opportunity to explore many 
physical processes in high energy astrophysics. These sources are ideal candidates for probing the jet physics, 
particle acceleration of UHECRs, intergalactic magnetic field (IGMF) and density of low energy background photons 
from CMB to optical wavelengths in the intergalactic space.
\begin{table}
\begin{tabular}{lclclclclclc}
\hline
Name	      &Redshift ($z$)	&$\Gamma_{3FGL}$ &$\Gamma_{3FHL}$ &$\Gamma_{TeV}$ &VHE Observation\\
\hline
RGB J0710+591 &0.125	        &1.661$\pm$0.094 &1.022$\pm$0.370 &2.64$\pm$0.27 &VERITAS \cite{Acciari2010}\\
1ES 0229+200  &0.140            &2.025$\pm$0.150 &1.541$\pm$0.401 &2.61$\pm$0.14 &VERITAS \cite{Aliu2014}\\
1ES 0347-121  &0.188            &1.734$\pm$0.156 &1.830$\pm$0.290 &3.09$\pm$0.20 &H.E.S.S. \cite{Aharonian2007}\\
1ES 0414+009  &0.287            &1.745$\pm$0.114 &1.851$\pm$0.337 &3.43$\pm$0.58 &VERITAS  \cite{Aliu2012}\\
\hline
\end{tabular}
\caption{Summary of the results from the $\gamma$-ray observations of the ESHP blazar candidates.}
\label{tab1}	
\end{table}
\section{EHSP Candidates}
The EHSP blazars are also known as hard spectra blazars.  Due to their hard $\gamma$-ray spectra, the EHSP blazars are 
faint in the energy range (0.1-500 GeV) covered by the Large Area Telescope (LAT) onboard the \emph{Fermi} satellite. 
However, a small group of this new and emerging class of blazars have been detected in GeV energy range by the 
\emph{Fermi}-LAT and at TeV energies by the current generation ground based Imaging Air Chererenov Telescopes (IACTs) like 
\emph{VERITAS, MAGIC and H.E.S.S.}. The results from the long term  \emph{Fermi}-LAT observations are reported in its different 
$\gamma$-ray catalogs. In this study, we have selected four well known EHSP candidate sources for which quasi-simultaneous 
observations from the \emph{Fermi}-LAT and IACTs are available. In the Third \emph{Fermi}-LAT source catalog (3FGL), 
time averaged spectral measurements in the energy range 0.1-100 GeV have been reported using four years of science data 
collected during the period August 2008- July 2012 \cite{Acero2015}. The Third Catalog of Hard \emph{Fermi}-LAT sources (3FHL)
provides results above 10 GeV from the seven years of observations between August 2008 to August 2015 \cite{Ajello2017}. This 
indicates that the \emph{Fermi}-LAT observations of the HE $\gamma$-ray sky provide an excellent overlap with the ground based 
VHE observations since its scientific operation began in 2008. The differential spectra of $\gamma$-ray emission from the EHSP 
blazars measured using \emph{Fermi}-LAT or IACTs are described by a power law of the form:
\begin{equation}\label{diff-spec}
	\frac{dN}{dE} = N_0 \left(\frac{E}{E_0}\right)^{-\Gamma}	
\end{equation}	
where $N_0$ is the normalization flux at the scale (pivot) energy $E_0$ and $\Gamma$ is the photon spectral index. 
The values of spectral indices reported in in the 3FGL ($\Gamma_{3FGL}$) and 3FHL ($\Gamma_{3FHL}$) along with the VHE 
observations ($\Gamma_{TeV}$) for four ESHP candidates selected in the present study are reported in Table \ref{tab1}. 
It is evident from Table \ref{tab1} that the observed TeV spectral indices are significantly softer than the \emph{Fermi}-LAT spectral 
indices in GeV energy band.
\begin{center}
\begin{figure}
\includegraphics[width=0.5\textwidth,angle=-90]{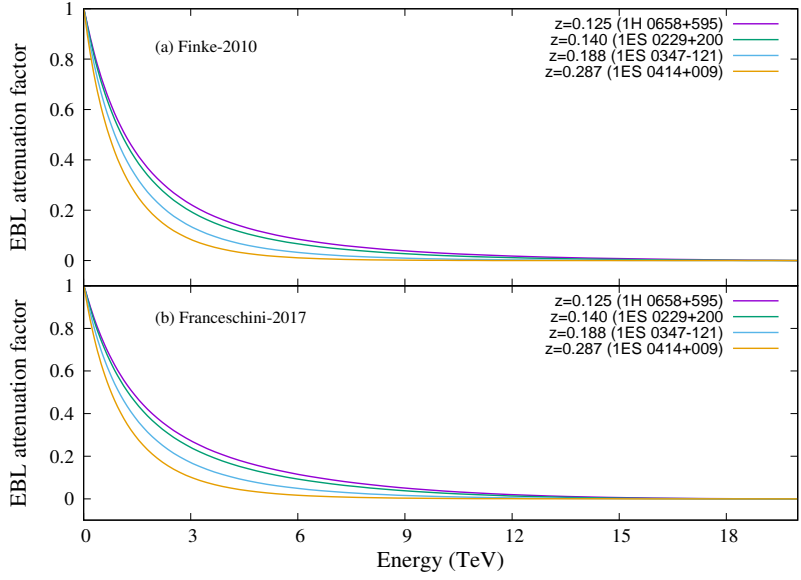}
\caption{Attenuation of $\gamma$-ray photons due to EBL absorption for sources at various redshifts using two different EBL models 
	proposed by (a) Finke et al. (2010) and (b) Franceschini et al. (2017).}
	\label{ebl-abs}
\end{figure}	
\end{center}	
\begin{figure}
\includegraphics[width=0.5\textwidth,angle=-90]{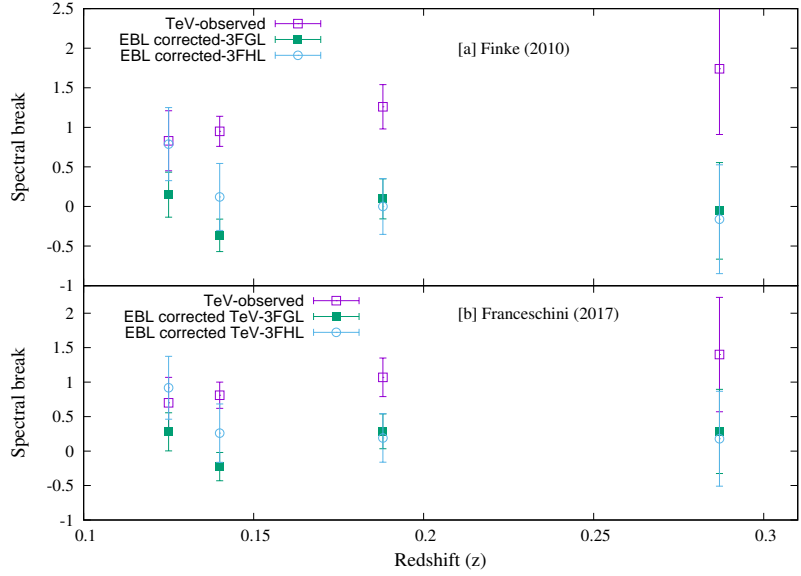}
	\caption{Spectral Break between $\Gamma_{int}$ and $\Gamma_{TeV}$, $\Gamma_{3FGL}$ \& $\Gamma_{3FHL}$  
	         for EHSP candidate blazars using (a) Finke et al. (2010) and (b) Franceschini et al. (2017) models.}
	\label{spec-brk}
\end{figure}	
\begin{figure}
\includegraphics[width=0.4\textwidth,angle=-90]{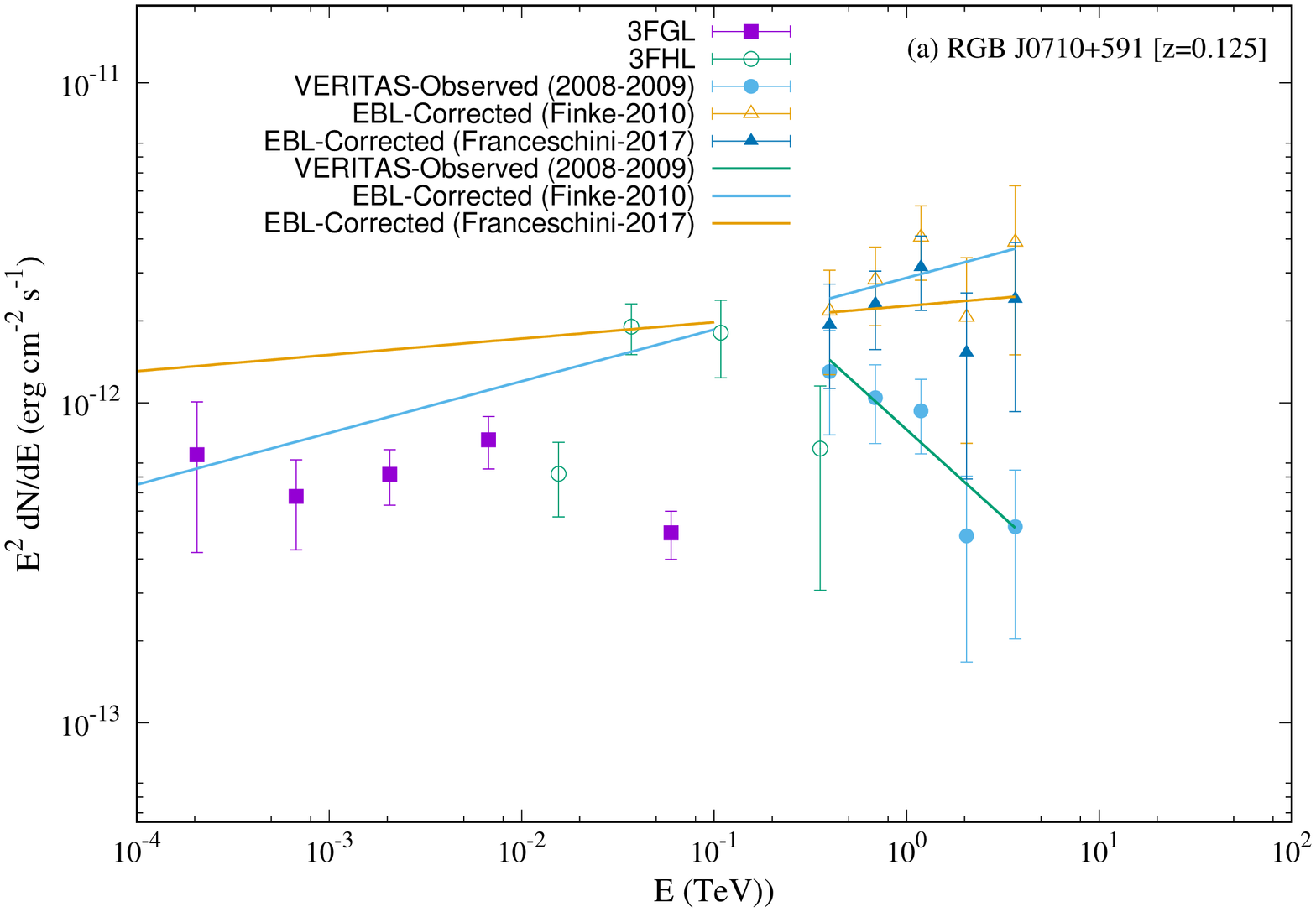}
\includegraphics[width=0.4\textwidth,angle=-90]{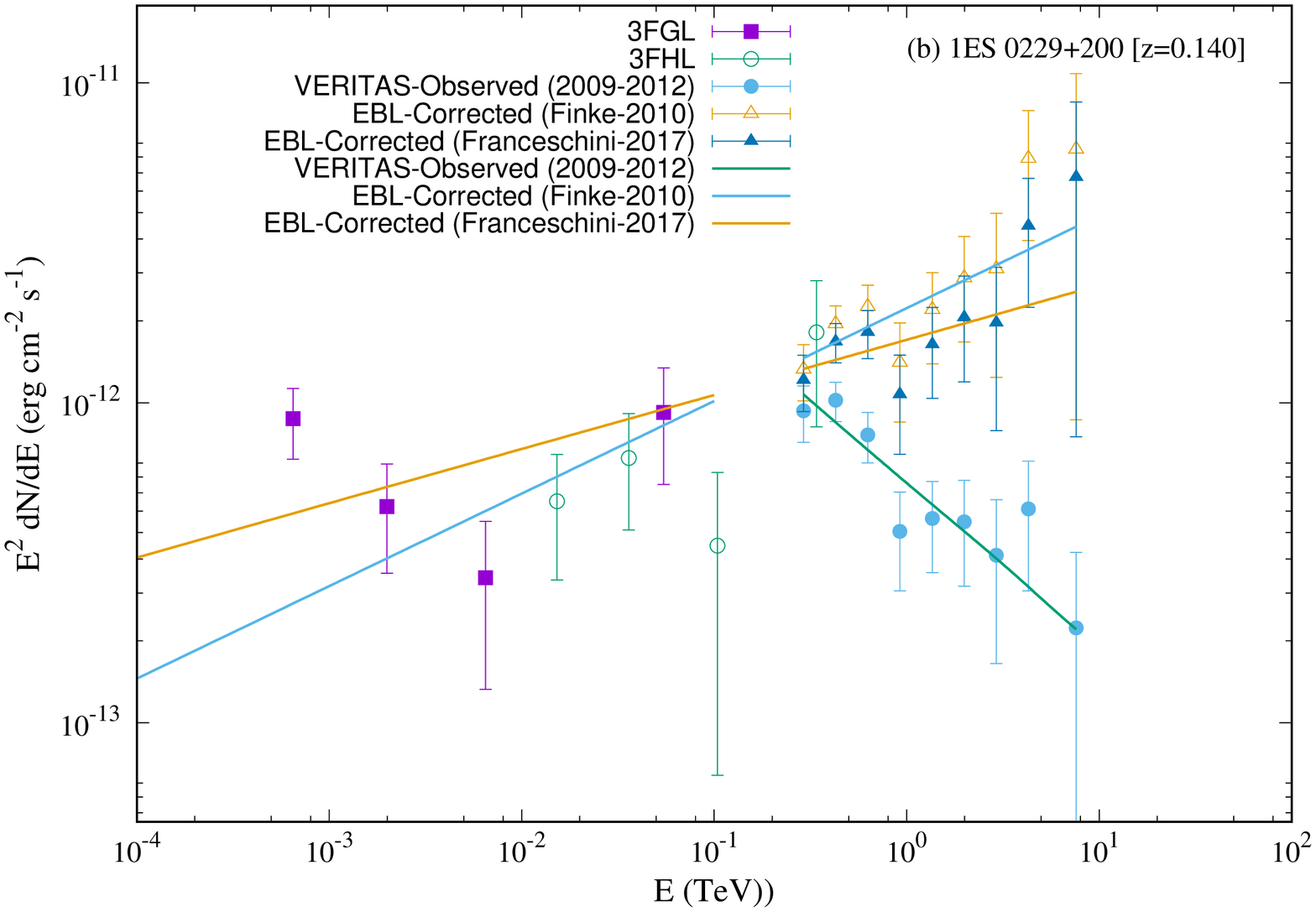}
\includegraphics[width=0.4\textwidth,angle=-90]{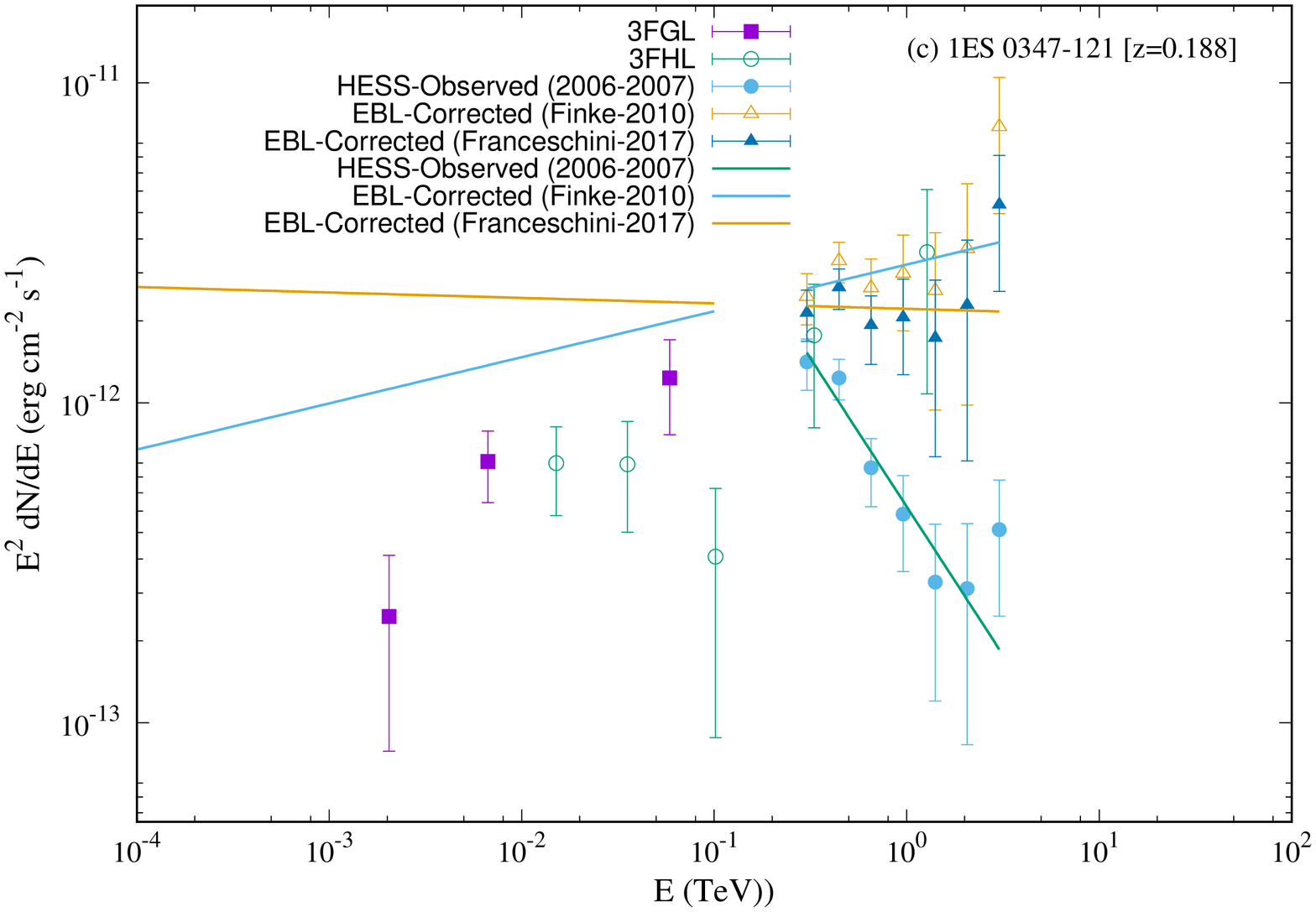}
\includegraphics[width=0.4\textwidth,angle=-90]{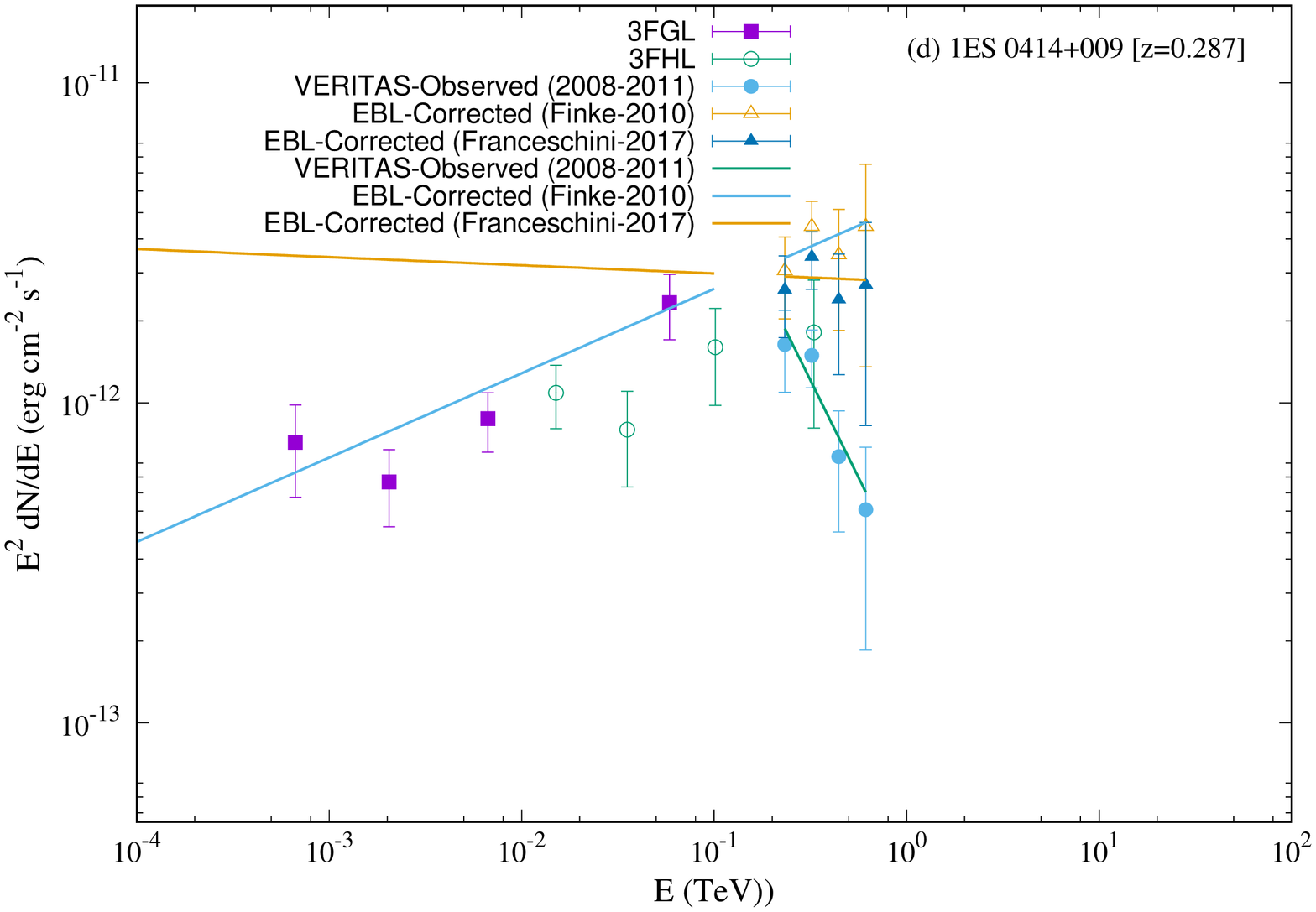}
\caption{GeV-TeV spectral energy distribution of EHSP blazars.}
\label{sed}
\end{figure}
\vspace{-1.2cm}
\section{Results and Discussion}
\subsection{Intrinsic VHE Spectrum}
The VHE $\gamma$-ray observations above 100 GeV by ground based telescopes suffer an opacity effect due 
to the presence of Extragalactic Background Light (EBL) in the intergalactic space. The EBL consists of 
IR-optical-UV photons emitted by the stars and galaxies throughout the evolution of the Universe. 
VHE $\gamma$-rays emitted from the cosmic sources are absorbed by the low energy (10$^{-3}$ eV - 10 eV) EBL 
photons via $\gamma-\gamma$ pair production. This process can partially remove the VHE photons from the 
beam propogating towards the observer and causes opacity to TeV observations of the sources in the 
extragalactic Universe. The opacity of the Universe is characterized by the \emph{optical-depth} ($\tau$) 
which strongly depends on the redshift of the source ($z$), energy of VHE $\gamma$-rays ($E$) and 
density of the EBL photons. The method of computing $\tau (E,z)$ for a given model of EBL photon density is 
described in \cite{Singh2014}. The VHE $\gamma$-ray spectrum of a source observed by the ground based instruments  
is modified with respect to the intrinsic spectrum and both are related as:  
\begin{equation}\label{int-spec}
	\left(\frac{dN}{dE}\right)_{obs} = \left(\frac{dN}{dE}\right)_{int}~~e^{-\tau (E, z)}
\end{equation}
where the term e$^{-\tau (E, z)}$ is commonly referred to as \emph{EBL attenuation factor}. 
The \emph{EBL attenuation factor} as a function of energy in the range 0.1 GeV to 20 TeV for four ESHP 
blazars at different $z$ is shown in Figure \ref{ebl-abs} (a) \& (b) corresponding to the two widely used 
EBL models proposed by Finke et al. (2010) \cite{Finke2010} and Franceschini et al. (2017) \cite{Frances2017} 
respectively. We observe that the attenuation due to EBL for both the models is negligible up to 100 GeV for 
all sources considered in the present study. However, TeV $\gamma$-ray photons above 100 GeV suffer large 
EBL absorption. This can lead to the significant softening of the intrinsic VHE spectra emitted from the sources 
at different redshifts.  We have derived the intrinsic TeV spectra for all the sources from the observed flux points 
after correcting for EBL absorption (Equation \ref{int-spec}) using two different models. The intrinsic TeV spectra 
are also described by a power law with spectral indices $\Gamma_{int}$ which are siginificantly harder than the 
observed spectral indices $\Gamma_{TeV}$ reported in Table \ref{tab1}. 
\subsection{GeV-TeV Spectral Break and Spectral Energy Distribution}
The absorption of VHE $\gamma$-rays due to interaction with EBL suggests that observed TeV spectrum of a source 
is different from the intrinsic TeV spectrum and the GeV spectrum measured from the \emph{Fermi}-LAT. The difference 
between any two power law spectral indices is referred to as \emph{spectral break}. We have estimated the spectral 
break caused by the absorption of TeV photons for the EHSP candidate blazars (Table \ref{tab1}) using two 
different EBL models and the results are shown in Figure \ref{spec-brk} (a) \& (b) as a function of redshift. 
It is found that the spectral break between observed and intrinsic TeV spectra ($\Gamma_{TeV}-\Gamma_{int}$) increases 
with redshift whereas the break between the \emph{Fermi}-LAT spectral indices ($\Gamma_{3FGL}~ \& ~\Gamma_{3FHL}$) and 
$\Gamma_{int}$ is nearly zero for all redshifts.
\par	
The SED of four EHSP blazars in the energy range 0.1 GeV to few TeV are shown in Figure \ref{sed}(a-d). The intrinsic TeV 
spectra of all the sources estimated using the EBL model proposed by Finke et al. 2010 \cite{Finke2010}, are described by a 
power law with an average photon spectral index $\Gamma_{int} \sim $ 1.7. This is broadly consitent with average of 
corresponding $\Gamma_{3FGL}$ and $\Gamma_{3FHL}$ values. The extrapolation of the intrinsic TeV spectrum following a power 
law in the GeV energy band suggests that the HE flux points measured by the \emph{Fermi}-LAT are generally below the flux 
level expected from the VHE observations. For 1ES 0347-121, the HE emission in the range 100 MeV to 1 GeV is 
below the detection sensitivity of the \emph{Fermi}-LAT and deviation of the flux points above 1 GeV from the TeV 
extraploation is also relatively large. The differential spectra of $\gamma$-ray photons with hard spectral index 
$\sim$ 1.7 suggest that the HE component of SED peaks at TeV energies. As discussed in Section 2, such hard TeV spectra can be 
mostly produced by the various electron distributions like hard power law, low-energy cutoff or Maxwellian with tail at 
higher energies. These spectral features in the electron distribution should also be observed in the synchrotron spectrum at 
low energies up to X-rays as synchrotron radiation is assumed to be produced by the same population of electrons in the 
frame-work of simple leptonic models. However, the synchrotron components in the broad-band SED of EHSP candidate blazars do 
not show such spectral features. This indicates that hadronic processes are the possible explanation for GeV-TeV emission from 
these sources. 
\vspace{-0.2cm}
\section{Conclusions}
The intrinsic TeV spectral indices of EHSP candidates are in good agreement with the \emph{Fermi}-LAT spectra. 
However, the flux points measured from the \emph{Fermi}-LAT are below the extrapolated TeV flux levels for all 
the sources considered in the present study.  The combined GeV-TeV $\gamma$-ray spectra of all the EHSP candidates 
are described by a single power law with photon spectral index $\Gamma~\sim$ 1.7 which is consistent with the 
proton synchrotron radiation due to injection spectral index of $\sim$ 2.4 in the emission region.
However, simultaneous long term monitoring of these sources in a wide GeV-TeV energy range by the upcoming 
Cherenkov Telescope Array (CTA)-observatory will be very important to fully explore the $\gamma$-ray emission processes 
from this new class of blazars. This will consequently help in constraining the EBL photon density permeated in the 
intergalactic space.

\end{document}